\documentclass[11pt,twoside]{article}
\usepackage{asp2006}
\usepackage{epsf}
\usepackage{psfig}
\usepackage{lscape}

\begin{document}
\title{Paper to Screen:  Processing Historical Scans in the ADS} 
\author{Donna M. Thompson, Alberto Accomazzi,Guenther Eichhorn,
  Carolyn Grant, Edwin Henneken, Michael J. Kurtz, Elizabeth Bohlen,
  Stephen S. Murray}   %%% Fill in author names
\affil{Harvard-Smithsonian Center for Astrophysics, 60 Garden Street,
  Cambridge, MA 02138}    %%% Fill in author affiliations 

\begin{abstract} %%% Abstract to run on from here.
The NASA Astrophysics Data System in conjunction with the Wolbach
Library at the Harvard-Smithsonian Center for Astrophysics is working
on a project to microfilm historical observatory publications.  The
microfilm is then scanned for inclusion in the ADS.  The ADS currently
contains over 700,000 scanned pages of volumes of historical literature.  Many
of these volumes lack clear pagination or other bibliographic data
that are necessary to take advantage of the searching capabilities of
the ADS.  This paper will address some of the interesting challenges
that needed to be resolved during the processing of the Observatory
Reports included in the ADS.   

\end{abstract}

%%% MAIN BODY OF TEXT GOES HERE. CONSULT "INSTRUCTIONS FOR AUTHORS USING
%%% LATEX2E MARKUP", SECTIONS 2.3-2.6 FOR HELP WITH EQUATIONS, FIGURES,
%%% AND TABLES.

\section*{Brief overview of the process of metadata capture}

%\subsection*{}   %%% Unnumbered second level section head (remove "%" symbol)

In order to be able to utilize the sophisticated searching
  capabilities of the ADS for the scanned 
  publications, page numbers and article metadata (e.g. title,
  author, beginning and ending pages) must be generated.  A software
  tool to aid in the capture of these metadata has been developed and
  is available on the ADS web site.

A number of volunteers have worked with this interface and have
generated the data for approximately 450 volumes of 44 different
titles.  Some of the observatory publications that have been processed
and are now searchable with the ADS include selections from the following:
{\it Annals of Harvard College Observatory}; {\it Astronomical and
  Meteorological Observations made at the U.S. Naval Observatory}; {\it
  Beobachtungs-Ergebnisse der Koniglichen Sternwarte zu Berlin}; and
{\it Contributions from the Rutherford Observatory of Columbia
  University New York}. 

The capture of the metadata must be done in two stages.  In the first
stage, page numbering mode, the scans are viewed sequentially and a page
number is assigned to each image and duplicate scans are marked to be
removed.  (See Figure 1.)  Once this process has been completed the
second stage, article metadata mode, can be done.  At this time the
images are shown with the assigned page numbers and the article
information (author, title, first/last page, and abstract) can be
entered.  (See Figure 2.)  Once these data have been checked and
processed further by the ADS staff the articles in this collection
become searchable using the ADS.   

\begin{figure}[!ht]
\plotone{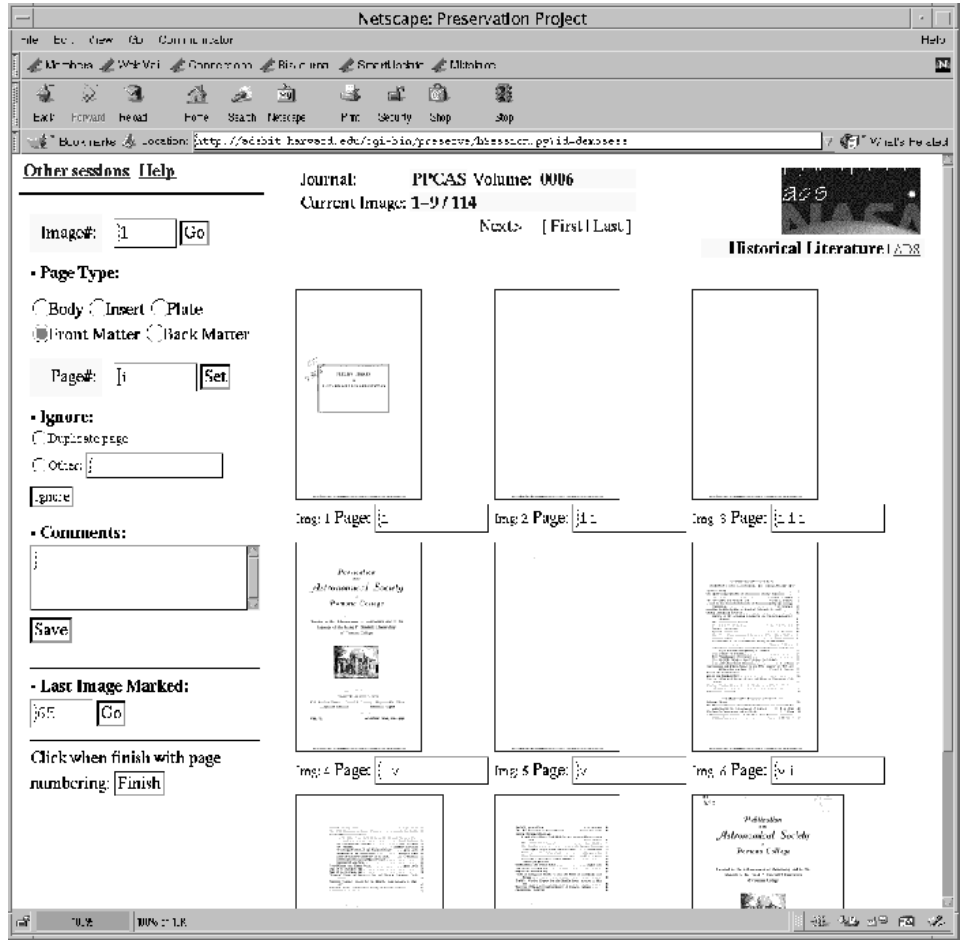}
\caption{Page Numbering Stage}
\end{figure}

\begin{figure}[!ht]
\plotone{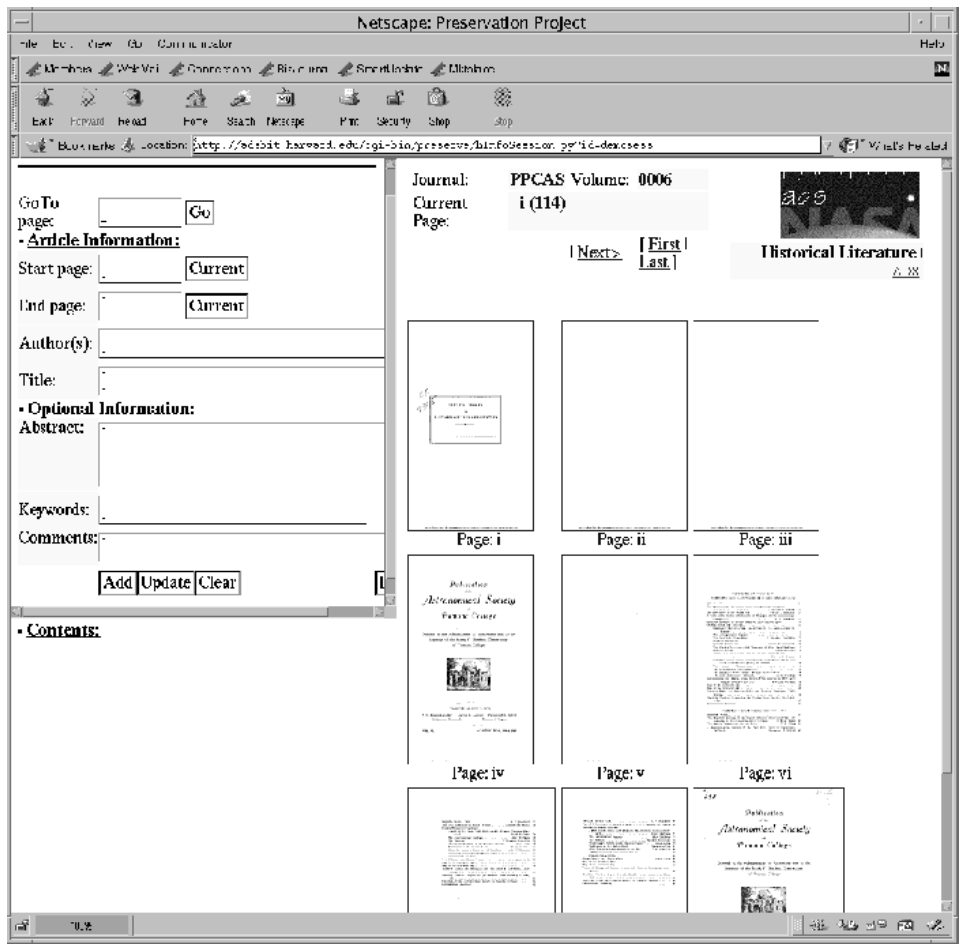}
\caption{Article Entering Stage}
\end{figure}

Many volunteers from around the world have participated in this
project.  As of June 2006 more than 120 people signed up as
contributors.  A Smithsonian Seidell grant was received and several library
school students were hired to work on the metadata capture, yet the
bulk of the work to date was done by several dedicated individuals.

\section*{What's in a Name?}
The ADS assigns a bibcode to each entry.  The bibcode is a 19
character string generated from the journal, volume and page
information for each publication. 
Many of the observatory publications required special handling to
create a usable bibcode.  
The form of the bibliographic code is:
 yyyyjjjjjvvvvmppppa
where: yyyy: year of publication; jjjjj: a standard abbreviation for
the journal; vvvv: the volume number (for a journal) or an abbreviation that
specifies what type of publication it is (e.g. conf for conference
proceedings); m: Qualifier for publication: e.g. L: Letter, Q-Z:
Unduplicating character for identical codes, letter designations on
pages; pppp: page number; a: the first letter of the last name of the
first author. 

We assign a journal abbreviation (bibstem) which is a 5 character long
string which identifies a publication.  And similar to cataloging
items for a print collection, we try to maintain continuity of a
publication when a name change occurs.  In some cases bound volumes were
broken up or merged into ``logical'' series with related bibstems in
order to be able to differentiate between the volumes.    

These decisions are made by checking bibliographic sources and by
checking the astronomical literature to see how the original sources
are referenced.  When a compelling argument can be made, new
abbreviations are added to the list.  Among the series which were
split:
\vskip 0.3cm
Annalen der Universitaets-Sternwarte Wien has been divided into:
AnWie and AnWiD Dritte Folge--third series
\vskip 0.3cm
Bulletin Astronomique has been divided into: BuAst--Bulletin
Astronomique; BuAsI--Bulletin Astronomique, Serie I; 
BuAsR--Bulletin Astronomique, Revue Generale des Travaux
Astronomiques.

These titles were divided into different series because leaving them
in one series created misleading or sometimes incorrect volume numbers
in the bibcode.  Adding new bibstems allows us to keep the bibcode
in alignment with what is written on the printed page. 

\section*{Dates and Different Page Numbers}
Another recurrent issue was that many of the Observatory Publications
were titled ``Reports of the XYZ Observatory for the year 19XX.''  The
year that we use in creating the bibcode is the publication year of the volume
which was frequently not the same year which the report was written
about.  To address this issue, we continue to use the publication year
in the bibcode but the report years are listed in the title field.  

An example is:  
\vskip 0.2cm
Title:  Reports for the Years 1900 to 1904, Presented by the Board of
Managers of the Observatory of Yale University to the President and Fellows
\vskip 0.1cm
Authors: Elkin, William L.
\vskip 0.1cm
Journal: Reports for the year presented by the Board of Managers of the
Observatory of Yale University to the President and Fellows, vol. 1,
pp.1.1-1.8 
\vskip 0.1cm
Publication Date: 00/1910
\vskip 0.1cm
Origin: ADS
\vskip 0.1cm
Bibliographic Code: 1910YalRY...1....1E
\vskip 0.2cm

Some publications used letters to designate different parts of the
reports and page numbers such as D4 were listed in the Table of
Contents.  To deal with these types of page numbering schemes we
needed to use our qualifier for publication field (see m in the
bibcode description.)  Hence ``Publications of the U.S. Naval
Observatory Second Series, vol. 4, pp. D:1-D:305'' is given a bibcode of
1906PUSNO...4D...1.  Pagination must ensure uniqueness of page number
within a volume, and attempts to reflect original (printed) page.

Another issue which periodically came up was that several
publications had handwritten numbers on the pages in addition to those
printed on the typed page.  These numbers were a result of changes
that were sometimes indicated in the editorial notes of various
publications.  Again in this situation, we investigated how these
volumes were cited in the literature and used the appropriate
numbering based on the results of this research.  

As a result of these oddities, it may be difficult or confusing for a
user to find a particular article by searching using the journal name,
volume number, page number or publication year. Author or title
searches in the historical literature would produce better results.

\acknowledgements %%% Text of acknowledgements runs on after this command.
The ADS is funded by NASA Grant NNG06GG68G.

%%% THE BIBLIOGRAPHY
%%%
%%% CONSULT SECTION 3 OF "INSTRUCTIONS FOR AUTHORS" FOR HOW TO USE NATBIB.
%%% AUTHORS ARE ENCOURAGED TO USE EITHER THE "THEBIBLIOGRAPY" ENVIRONMENT
%%% BY UNCOMMENTING (DELETING THE "%" SYMBOL) THE COMMANDS BELOW, OR BY
%%% USING THE BIBTEX ENVIRONMENT. TO FIND OUT WHICH IS APPLICABLE TO YOUR
%%% CONTRIBUTION, CONSULT THE VOLUME EDITORS FOR YOUR PROCEEDINGS.
%%%

%\begin{thebibliography}{}
%\bibitem[]{}
%\bibitem[]{}
%\bibitem[]{}
%\bibitem[]{}
%\bibitem[]{}
%\bibitem[]{}
%\bibitem[]{}
%\bibitem[]{}
%\bibitem[]{}
%\bibitem[]{}
%\bibitem[]{}
%\bibitem[]{}
%\end{thebibliography}

\end{document}